# Temperature and Electron Density Dependence of Spin Relaxation in GaAs/AlGaAs Quantum Well

L. F. Han · Y. G. Zhu · X. H. Zhang* · P. H. Tan · H. Q. Ni · Z. C. Niu

State Key Laboratory for Superlattices and Microstructures, Institute of Semiconductors, Chinese Academy of Sciences, P.O. Box 912, Beijing 100083, People's Republic of China

**Abstract:** Temperature and carrier density dependent spin dynamics for GaAs/AlGaAs quantum wells (QWs) with different structural symmetry has been studied by using time-resolved Kerr rotation technique. The spin relaxation time is measured to be much longer for the symmetrically-designed GaAs quantum well comparing with the asymmetrical one, indicating the strong influence of Rashba spin-orbit coupling on spin relaxation. D'yakonov-Perel' (DP) mechanism has been revealed to be the dominant contribution for spin relaxation in GaAs/AlGaAs QWs. The spin relaxation time exhibits non-monotonic dependent behavior on both temperature and photo-excited carrier density, revealing the important role of non-monotonic temperature and density dependence of electron-electron Coulomb scattering. Our experimental observations demonstrate good agreement with recently developed spin relaxation theory based on microscopic kinetic spin Bloch equation approach.

---

*Corresponding author. E-mail address: xinhuiz@semi.ac.cn.
Tel:+82-010-82304870; Fax: +82-010-82305056

# Introduction

Spin dynamics and the related physics in semiconductors have drawn much attention in the past years for its importance to realize novel spin-electronic devices [1]. In recent years, electron spin relaxation in many types of materials, especially in low dimensional III-V group semiconductor heterostructures, has been studied extensively both theoretically and experimentally [1]. The relevant spin relaxation mechanisms, such as the Elliott-Yafet (EY), Bir-Aranov-Pikus (BAP) and D'yakonov-Perel' (DP) mechanisms as well as hyperfine interactions, have been well established to describe spin relaxation and dephasing dynamics. However, it is known that the relative importance of these mechanisms strongly depends on material design and temperature as well as carrier concentration so on. Previous investigations in literature show that the BAP mechanism dominates the spin relaxation at low temperatures for bulk GaAs [2,3] and GaAs/AlGaAs quantum wells (QWs) [4,5], whereas DP mechanism dominates spin relaxation in other regimes. However, recent reexaminations using the microscopic kinetic spin Bloch equation approach [6-9] have revealed that BAP mechanism is much less important than DP mechanism for intrinsic III-V group semiconductors, even at low temperatures. The DP mechanism resulting from spin–orbit coupling in systems lacking inversion symmetry (such as zinc-blende structure or asymmetric confining potentials in QWs), has a spin relaxation rate inversely related to the momentum scattering rate [9]. Electron spin relaxation in GaAs QWs has been experimentally studied through temperature [10, 11] and QW width dependence [10–12], and DP mechanism has been revealed to dominate spin relaxation in intrinsic QWs at high temperatures [10]. The oscillatory spin dynamics study for two-dimensional electron gas (2DEG) at low temperatures also demonstrated the dominance of DP mechanism in the weak momentum scattering regime [13]. The observed enhancement of spin relaxation time resulting from electron-electron scattering in *n*-doped GaAs/AlGaAs QW agrees with DP mechanism governed by electron-electron scattering as well [14-17]. The experimental observation of electron spin relaxation time maximum for temperature dependent study in a high-mobility GaAs/AlGaAs 2DEG has also revealed the importance of electron-electron Coulomb scattering [18].

It is well known that the spin-orbital (SO) coupling leads to a strong momentum dependent mixing of spin and orbital-momentum eigenstates, so that scattering processes change spin and orbital angular momentum, therefore contribute to spin relaxation accordingly [19-21]. For electrons in two-dimensional semiconductor heterostructures or quantum wells, the Rashba spin-orbit coupling due to structure inversion asymmetry and the Dresselhaus spin-orbit coupling due to bulk inversion asymmetry in the compounds cause electron spin relaxation and decoherence through spin precession of carriers with finite crystal momentum ***k*** in the effective ***k***-dependent crystal magnetic field of an inversion-asymmetric material. Therefore, spin relaxation and decoherence studies in semiconductors have revealed important physics of spin-orbital coupling. Since carrier spin relaxation is related to several competing mechanisms and particularly different material and structure design, in which different spin-orbital coupling is involved to spin relaxation processes. Thus the experimental investigation of spin relaxation and its dependence on temperature and carrier density has been found to vary widely between different samples. In this letter, we have designed two GaAs/AlGaAs quantum wells with different structural symmetry. The spin relaxation time is measured to be much shorter for the asymmetrically-designed GaAs/AlGaAs QW comparing with the symmetrical one, indicating the strong effect of Rashba spin-orbit coupling on spin relaxation. The comprehensive studies of temperature and carrier density dependence of spin relaxation time for both samples have revealed that electron spin relaxation in GaAs/AlGaAs QWs is governed mainly by DP mechanism in the entire temperature regime. The spin relaxation time exhibits non-monotonic behavior for both temperature and photo-excited carrier density dependence, revealing the important role of non-monotonic temperature and density dependence of electron-electron Coulomb scattering. Our experimental observations demonstrate good agreement with recently developed spin relaxation theory based on microscopic kinetic spin Bloch equation approach [6-9].

**Experimental Details**

The samples used in our experiments were *n*-modulation doped single GaAs/AlGaAs quantum well with well width of 10 nm grown by molecular beam epitaxy on (100)-oriented semi-insulating GaAs substrates. As shown in Fig. 1, one of the samples is designed to be symmetric with two-side symmetric $Al_{0.3}Ga_{0.7}As$ barriers and δ-doping concentration of $2 \times 10^{10} cm^{-2}$. The other sample is designed to be asymmetric with different aluminum components at two sides ($Al_{0.3}Ga_{0.7}As$ for the up-side barrier and $Al_{0.25}Ga_{0.75}As$ for the down-side barrier) and asymmetric doping concentration as well (the nominal up-side *n*-modulation doping is $2\times10^{11}cm^{-2}$ and the down-side doping is $2\times10^{10}cm^{-2}$). The rest layer structures and growth conditions are identical for two samples.

In our time-resolved magneto-Kerr rotation (TR-MOKE) measurement, a Ti: Sapphire laser system (Coherent Chameleon Ultra II) provided 150 fs pulses with repetition rate of 80 MHz. The pump beam with central wavelength ranging from 770 to 860 nm was incident normal to the sample, while probe beam was at an angle of about $30^o$ to the surface normal. The polarization of the pump beam was adjusted to be circularly polarized and the probe beam was linearly polarized. The sample was mounted within a Janis closed-cycle optical cryostat which is located in between two poles of an electromagnet. After reflection from sample, the Kerr rotation signal was detected by a sensitive optical bridge and lock-in amplifier. The photoluminescence (PL) measurements have been first performed at wide temperature range to check sample's quality and identify the band-edge energies for the specially designed samples.

## Results and Discussions

Both samples have shown very good PL response up to room temperature as presented in Fig.2, indicating good sample quality. The PL peak at 848nm is related to the band edge of quantum well and the peak at 870nm is related to the band edge of

GaAs substrate. The band edge related PL peak for quantum well has blue-shifted to 798nm at 10K as expected. The time-resolved Kerr rotation measurements are then performed at different temperatures and photo-excitation powers with excitation wavelength centered either at the resonant position of the band edge PL peak of quantum wells or well above it. The typical time-resolved Kerr rotation signals measured at 20K, 200K and 250K with excitation wavelength of 798nm and pumping electron densities of $1.15 \times 10^{11} cm^{-2}$ for the asymmetric GaAs QW have been presented in Fig. 3 (a). The temperature dependence of electron spin relaxation times extracted from the exponential decay function fitting of time-resolved Kerr response for both symmetric and asymmetric GaAs/AlGaAs QWs is shown in Fig. 3 (b). It is seen that the measured spin relaxation time for the symmetric GaAs/AlGaAs QW is generally longer than the asymmetric one at low temperature regime (<100K). This result is consistent with the well-established DP spin relaxation mechanism, in which the spin-orbital coupling plays important role for electron spin relaxation. It is naturally expected that the Rashba spin-orbital coupling in the asymmetric sample should be stronger than the symmetric one, whereas the Dresselhaus spin-orbital coupling could be similar for two samples. The shorter spin relaxation time for the asymmetric QW is caused by much stronger inhomogeneous magnetic fluctuations resulting from stronger Rashba spin-orbital coupling than the symmetric QW. Meanwhile, electron spin relaxation times for both symmetric and asymmetric GaAs/AlGaAs QWs have shown non-monotonical dependence on temperature. Spin relaxation time increases first with increasing temperature, shows a peak value of 399 ps around 180 K, and then decreases with further increase of temperature for the asymmetric QW. The symmetric GaAs/Al$_{0.3}$Ga$_{0.7}$As QW shows the similar temperature dependent spin relaxation time behavior, but with a maximum spin relaxation time of 592 ps observed around 60K. The BAP mechanism, in which the spin relaxation is caused by spin-flip via the electron-hole exchange interaction, predicts that the spin-relaxation time decreases rapidly with increasing temperature in low temperature regime. However, the observed low temperature dependence of spin relaxation is in contrast to the BAP mechanism. Rather, the appearance of a peak in the temperature dependent spin relaxation time for both symmetric and asymmetric GaAs QW structures agrees well with the theoretical reexamination of spin dynamics in both quantum well and bulk intrinsic III-V group semiconductors [6-9], in which BAP mechanism has been shown to be much less important than DP mechanism, even at low temperatures. The strengthening momentum scatterings at higher temperatures suppress the inhomogeneous broadening (the random spin precession), and tend to prolong the spin relaxation time with DP mechanism, in which spin relaxation time is inversely proportional to the momentum scattering time $\tau_p$. The appearance of the peaks in Fig.3 (b) originates from the DP mechanism controlled by electron-electron Coulomb scattering. Moreover, as discussed in references 6-9, electron-electron Coulomb scattering rate $\tau_p^{ee}$ is a non-monotonic function of temperature and electron density with a minimum of $\tau_p^{ee}$ corresponding to the crossover from the degenerate limit to the non-degenerate one at Fermi temperature $T_F = E_F/K_B$ [6-9].

When temperature further increases, electron-phonon scattering will then strengthen and become comparable to electron-electron scattering, eventually dominate the spin relaxation process, thus spin relaxation time decreases with further rising temperatures. As a result, spin relaxation time shows a maximum. Considering the total electron density $n_e$ is the sum of optically excited carrier density and doping density (assuming fully ionized Si doping), the peak of temperature dependent spin relaxation time is calculated to appear at about 59K and 140K for the symmetric and asymmetric GaAs/AlGaAs QW samples under optically-pumped electron density of $1.15 \times 10^{11} \text{cm}^{-2}$, respectively. This, however, only agrees with the observed peak position for the symmetric sample. The inconsistence for the asymmetric sample may result from the uncertainty of the actual electron density.

The typical time-resolved Kerr rotation signals measured at 10K with excitation wavelength of 798nm at three different pumping electron densities for the asymmetric GaAs QW have been presented in Fig. 4 (a). The optically-pumped electron density dependence of electron spin relaxation times measured at 10K with excitation wavelength of 798nm for both symmetric and asymmetric GaAs/AlGaAs QWs is shown in Fig. 4 (b). Again, the electron spin relaxation times for both symmetric and asymmetric GaAs/AlGaAs QWs have shown non-monotonic dependence on excitation density with a peak appearing at electron density of $n_c = 1.5 - 1.7 \times 10^{11} \text{cm}^{-2}$. The non-monotonic density dependence of spin relaxation time mainly results from electron-electron Coulomb scatterings contribution at low temperature. It is known that, in the strong scattering regime, spin relaxation time can be expressed as $\tau_{s,i}^{-1} = \langle \Omega_\perp^2 \rangle \tau_p^*$ [6-9], where $\langle \Omega_\perp^2 \rangle$ is the average square of the component for precession vector in the plane perpendicular to $i$, and $\tau_p^*$ is the momentum scattering time. In the non-degenerate (low density) regime the electron-electron scattering increases with electron density, whereas the inhomogeneous broadening barely changes since the distribution function is close to the Boltzmann distribution. The spin relaxation is thus governed by strengthening electron-electron scattering with increasing carrier density. Therefore, spin relaxation time increases with increasing excitation density. In degenerate (high density) regime, both $\tau_p^{ee}$ and the inhomogeneous broadening $\langle \Omega_\perp^2 \rangle$ increases with electron density, leading to the decreased spin relaxation time with carrier density. The similar non-monotonic dependent behavior on excitation density for spin relaxation time has also been experimentally observed for intrinsic GaAs quantum well [22] and bulk CdTe crystal [23, 24] at room temperature. DP mechanism has been revealed to dominate the electron spin relaxation at room temperature. The doping density dependent electron spin relaxation investigation at low temperature for *n*-type bulk GaAs has also observed a peak in spin-dephasing times[25], and this peak has been attributed to the influence of electron screening and scattering on the spin dynamics of the excited electrons [25,26]. The high temperature regime (280-400K) investigation of spin relaxation time as function of carrier density for bulk GaAs also confirmed the dominance of DP and electron optical-phonon scattering mechanism [27]. Our experimental investigations for temperature and carrier density dependent electron spin dynamics give further demonstration of DP mechanism-governed spin relaxation process in *n*-modulation-doped GaAs QWs.

## Conclusions

In conclusion, the temperature and carrier density dependent studies of spin relaxation time for modulation-doped GaAs/AlGaAs quantum wells have demonstrated good agreement with recently developed spin relaxation theory based on microscopic kinetic spin Bloch equation approach. The spin relaxation time is measured to be much longer for the symmetrically-designed GaAs quantum well comparing with the asymmetrical one, indicating the strong influence of Rashba spin-orbit coupling on spin relaxation. DP mechanism has been revealed to dominate spin relaxation for *n*-modulation-doped GaAs QWs in the entire temperature regime. Our experimental results provide further fundamental understanding of spin dynamics in modulation-doped heterostructures towards potential semiconductor spintronics application based on GaAs/AlGaAs material systems.

## Acknowledgements:

This work is supported by the National Basic Research Program of China under Nos. 2011CB922200 and 2007CB924904; the National Natural Science Foundation of China under Nos. 10974195 and 10734060.

**Figure captions:**

Figure 1: Schematic diagram of the symmetric (left) and asymmetric (right) GaAs/AlGaAs quantum well structures grown by MBE.

Figure 2: The photoluminescence response measured at room temperature for both symmetric and asymmetric GaAs /AlGaAs quantum wells.

Figure 3: (a) Time-resolved Kerr rotation signals measured at 20K, 200K and 250K with excitation wavelength of 798nm and optically-pumped electron densities of $1.15 \times 10^{11} cm^{-2}$ for the asymmetric GaAs/AlGaAs quantum well. (b) Temperature dependence of electron spin relaxation times measured at photo-excitation electron density of $1.15 \times 10^{11} cm^{-2}$ for both symmetric and asymmetric GaAs/AlGaAs quantum well. The solid lines are drawn to guide eyes.

Figure 4: (a) The typical time-resolved Kerr rotation signals measured at 10K with excitation wavelength of 798nm at three different pumping electron densities for the asymmetric GaAs QW. (b) The photo-excited carrier density dependence of electron spin relaxation times measured at 10K for both symmetric and asymmetric GaAs /AlGaAs quantum well. The solid lines are drawn to guide eyes.

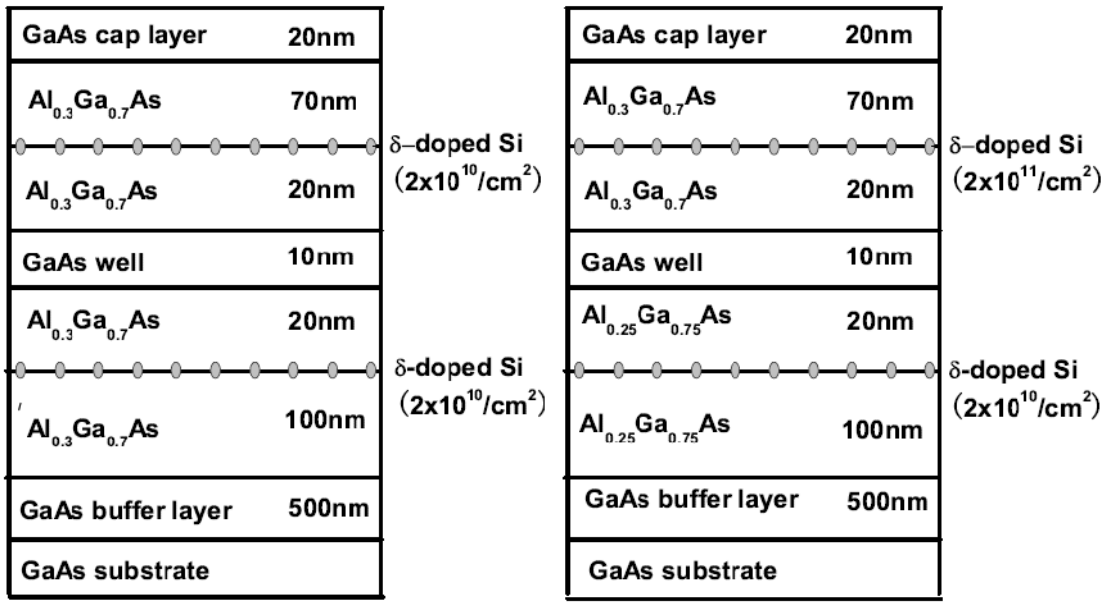

Figure 1

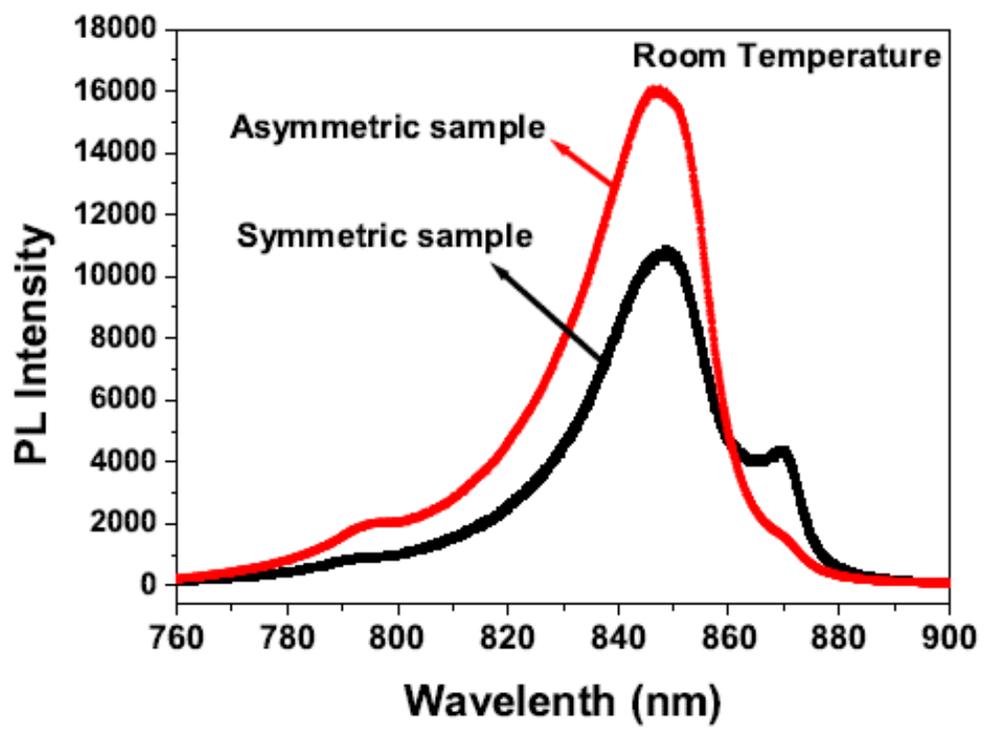

Figure 2

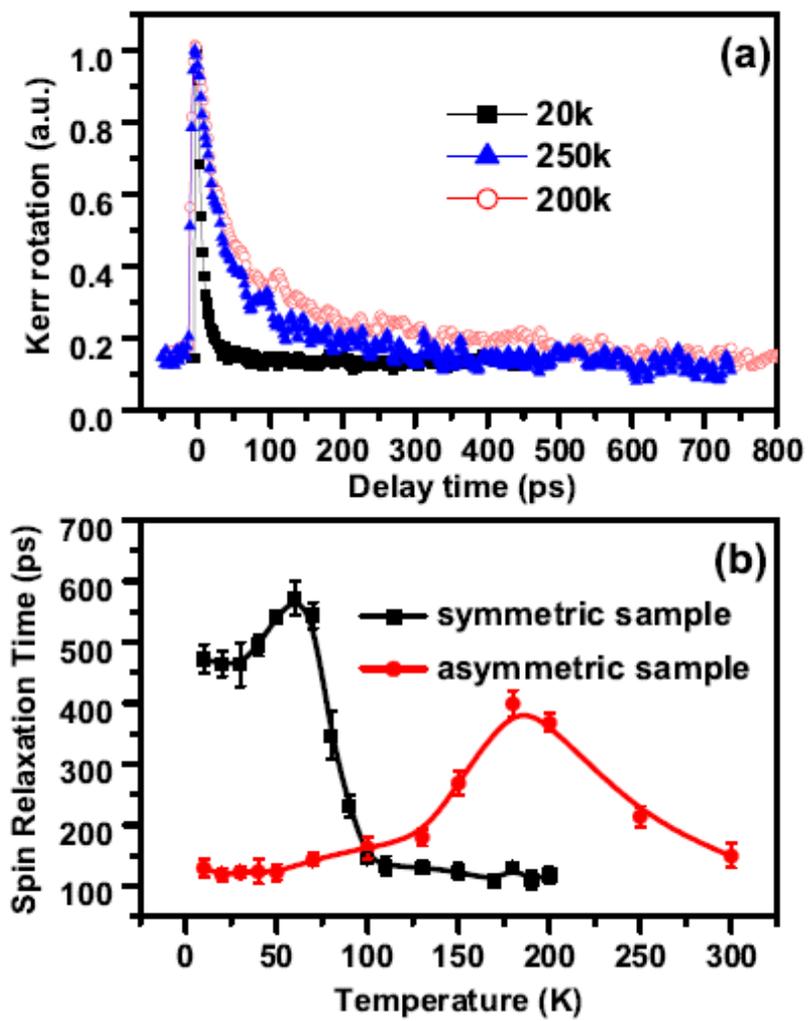

Figure 3

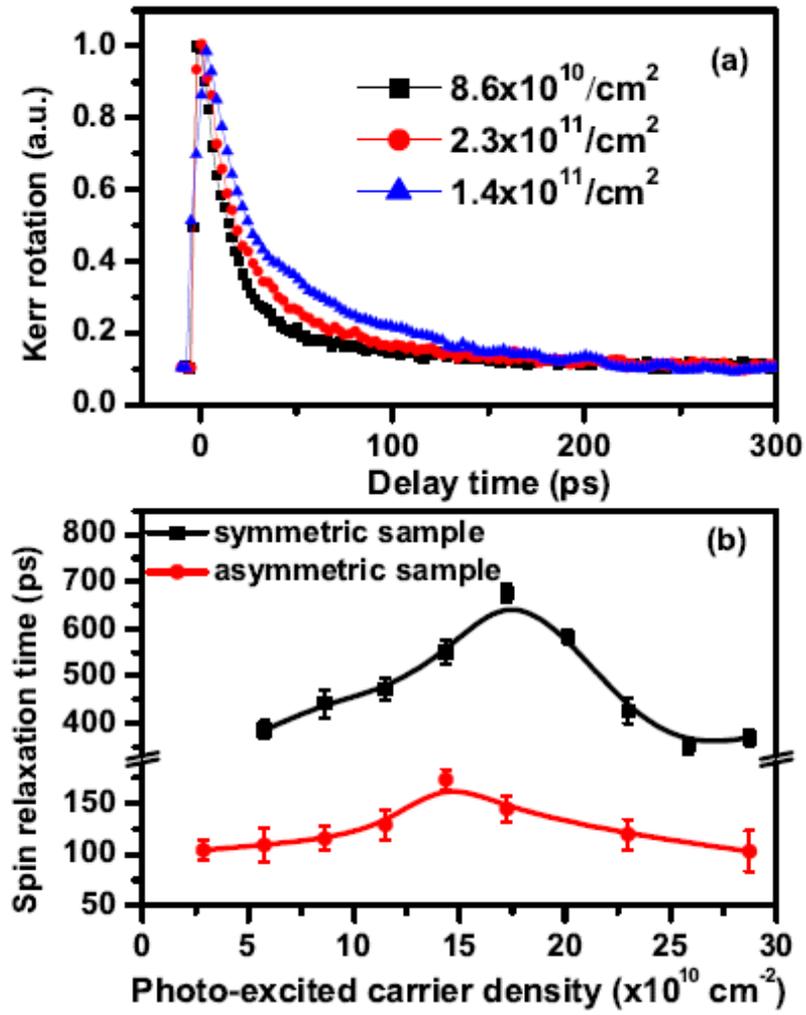

Figure 4